\documentclass[10pt,aps,twocolumn,
superscriptaddress,
footinbib,
noeprint,
longbibliography,
prb]{revtex4-2}

\usepackage{amsmath}
\usepackage{listings}
\usepackage{graphicx}% Include figure files
\usepackage{dcolumn}% Align table columns on decimal point
\usepackage{bm}% bold math
\usepackage{graphics}
\usepackage{bbold}
\usepackage{epsfig,color}
\usepackage{dsfont}
\usepackage{mathtools,amssymb}
\usepackage{braket}
\usepackage{stmaryrd}
\usepackage{bbm}

\usepackage{soul}

\usepackage{tikz}
\usetikzlibrary{arrows.meta,positioning}

\usepackage[breaklinks,colorlinks,bookmarks=false,citecolor=red,linkcolor=blue,urlcolor=magenta]{hyperref}

\begin{document}
\title{Graph Neural Networks for Fast Operator Selection in Adaptive VQE}

\author{ Javad Vahedi}
\email[]{javahedi@gmail.com}
\affiliation{Department of Physics and Earth Sciences and 
Department of Computer Science, Constructor University, Bremen 28759, Germany}

\author{Hadi H. Arefi,}
\email[]{hadi.arefi@aci.uni-hannover.de }
\affiliation{Leibniz University Hannover, Institute of Inorganic Chemistry, 30167 Hannover, Germany}
%\affiliation{Peter Grünberg Institut (PGI-3), Forschungszentrum Jülich, 52425, Jülich, Germany}

\date{\today} 

\begin{abstract}
Adaptive variational quantum algorithms construct problem-tailored ansätze by iteratively selecting operators from a predefined pool, typically using gradient-based criteria as in ADAPT-VQE. Although this strategy improves expressivity while avoiding unnecessarily large parameter spaces, it requires repeated scans over the full operator pool, so the classical cost of each iteration scales linearly with pool size. For systems with long-range interactions or large operator sets, this overhead can dominate the overall runtime. Here, we reformulate adaptive operator selection as a graph-based decision problem and introduce a graph neural network (GNN) policy that predicts the next entangling operator directly from the interaction graph and state-dependent observables. Training data are generated from exact simulations of disordered long-range spin chains, using gradient magnitudes, $|i\langle\psi|[H,P_{ij}]|\psi\rangle|$, as supervision signals. The learned policy accurately reproduces the dominant structure of the greedy gradient-based selection rule and significantly outperforms heuristic baselines based only on interaction strength. Integrated into a variational quantum eigensolver workflow, the resulting GNN-VQE approach achieves energy errors close to those of gradient-based ADAPT-VQE while reducing the need for repeated full-pool gradient evaluations during operator selection. To probe transfer beyond disordered spin models, we additionally construct small active-space molecular benchmarks for LiH and BeH$_2$ and find that the learned GNN is most effective as a shortlist generator: exact rescoring over only a few GNN-proposed candidates recovers near-oracle or oracle-level rollout behavior while using only a small fraction of the full candidate search. These results show that adaptive circuit construction contains learnable structure that can be exploited to reduce the classical cost of variational quantum algorithms.
\end{abstract}

\maketitle

\section{Introduction}
Variational quantum algorithms have emerged as one of the most promising approaches for studying quantum many-body systems on near-term quantum
devices~\cite{Peruzzo2014VQE,McClean2016Theory,Cerezo2021VQEReview}. Among these methods, the Variational Quantum Eigensolver (VQE) provides a hybrid quantum–classical framework for approximating ground states of
interacting Hamiltonians. In VQE a parametrized quantum circuit prepares a trial state
\begin{equation}
|\psi(\boldsymbol{\theta})\rangle = U(\boldsymbol{\theta})|0\rangle ,
\end{equation}
and the parameters $\boldsymbol{\theta}$ are optimized using classical
routines in order to minimize the energy expectation value
\begin{equation}
E(\boldsymbol{\theta}) =
\langle \psi(\boldsymbol{\theta}) | H | \psi(\boldsymbol{\theta}) \rangle .
\end{equation}

This hybrid optimization strategy has been widely explored in quantum chemistry and condensed matter physics as a method for approximating ground states using shallow quantum circuits on noisy quantum devices, and has become a central framework in contemporary discussions of quantum computing for chemistry and materials applications~\cite{Tilly2022,McArdle2020,Blunt2022,Evangelista2023, Alexeev2025}. In chemistry, this has motivated a broad effort to identify circuit constructions and active-space reductions that preserve chemically relevant correlation while remaining compatible with noisy intermediate-scale quantum hardware~\cite{deGraciaTrivino2023,Stein2016}.

A central challenge in variational quantum algorithms lies in the design of the circuit ansatz. The expressive power of the variational state
depends critically on the chosen gate structure, yet overly expressive circuits introduce large parameter spaces and optimization difficulties
such as barren plateaus~\cite{McClean2018Barren}. To address this issue, adaptive circuit construction methods have been
proposed in which the variational circuit is grown iteratively during the
optimization procedure.
One prominent example is the ADAPT-VQE algorithm~\cite{Grimsley2019ADAPT,Tang2021QubitADAPT}, where the ansatz is constructed from a predefined pool of operators. At each iteration the algorithm evaluates a gradient-like quantity
\begin{equation}
g_k =
\left|
\langle \psi | [H,O_k] | \psi \rangle
\right|
\end{equation}
for every operator $O_k$ in the pool, and the operator with the largest gradient magnitude is appended to the circuit. Related adaptive strategies have also been explored in other ansatz
construction schemes~\cite{Claudino2020, Yordanov2021, Lan2022, Feniou2023, Anastasiou2024, Vaquero-Sabater2024, Mullinax2025, Sun2025, Singh2025}.

Although this strategy systematically constructs expressive ansätze, its computational cost can become significant. If the operator pool contains $|E|$ candidate operators and the adaptive circuit grows to depth $T$, the algorithm requires $\mathcal{O}(T|E|)$ gradient evaluations. Each iteration therefore requires scanning the full operator pool to identify the most relevant operator. More broadly, practical VQE performance is often limited not only by ansatz quality but also by classical and measurement-side overheads associated with optimization, operator evaluation, and resource allocation~\cite{Li2024,Zhu2024}.

This challenge becomes particularly relevant in disordered quantum spin
systems with long-range couplings~\cite{Koffel2012,Vodola2014}. In such systems the interaction graph contains many possible two-site operators whose importance depends sensitively on the disorder
realization and the interaction-range exponent $\alpha$. Strongly disordered spin systems exhibit hierarchical structures that are well captured by strong-disorder renormalization group (SDRG) approaches~\cite{Fisher1994,Refael2013,Mohdeb2020,Mohdeb2022,Mohdeb2023}. In these methods, the strongest local interactions determine the formation of entangled clusters and guide the renormalization flow. This suggests that the selection of entangling operations in a variational circuit may follow structured, graph-dependent decision rules.

\begin{figure*}[t]
\centering
\includegraphics[width=\textwidth]{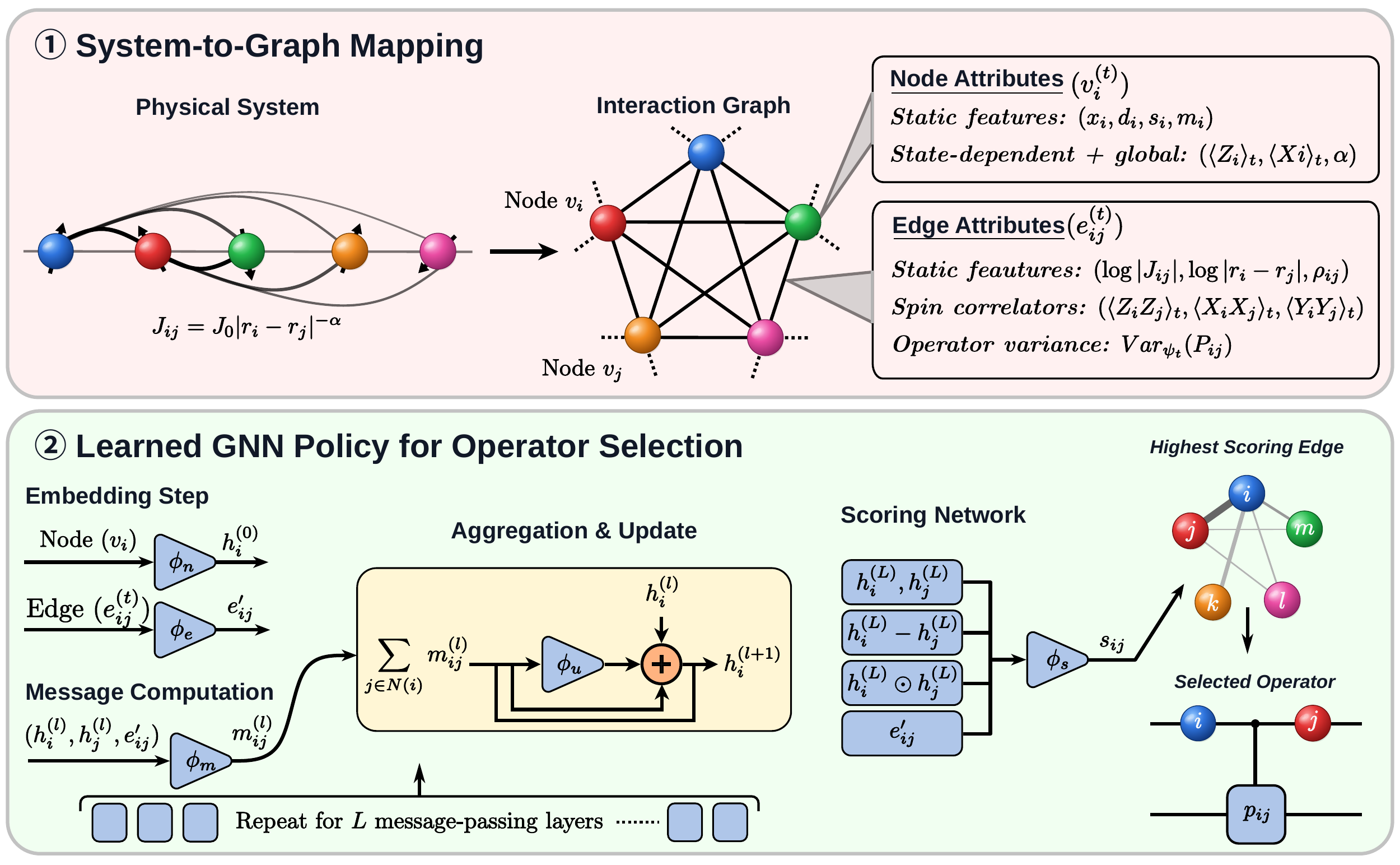}
\caption{
Schematic of the graph-based representation and learned GNN policy used for adaptive operator selection in the long-range spin-system setting studied in this work. \textbf{1. System-to-graph mapping:} The disordered spin chain is represented as an interaction graph in which nodes $v_i^{(t)}$ correspond to spins and edges $e_{ij}^{(t)}$ correspond to candidate two-qubit generators $P_{ij}$. Node features combine static geometric quantities with state-dependent observables, while edge features combine static interaction information, two-spin correlators, and operator-variance information. \textbf{2. Learned GNN policy for operator selection:} Node and edge features are first embedded into latent representations using $\phi_n$ and $\phi_e$. Edge-conditioned messages $m_{ij}^{(\ell)}$ are then computed with $\phi_m$, aggregated over neighbors, and used to update node embeddings through $\phi_u$ with a residual connection. After $L$ message-passing layers, candidate edges are scored by $\phi_s$ using the endpoint embeddings, their difference, their element-wise product, and the encoded edge feature $e'_{ij}$. The highest-scoring edge determines the next entangling operator selected during variational circuit construction.
}
\label{fig:gnn_architecture_map}
\end{figure*}

Graph neural networks (GNNs) provide a natural framework for learning such structured decision policies. By representing a quantum system as a graph whose nodes correspond to
spins and whose edges encode interaction strengths and local
correlations, message-passing architectures~\cite{Gilmer2017,Battaglia2018,Singh2022} can learn representations that capture both local connectivity and
global context. Graph-based machine learning has been successfully applied to a variety
of problems in physics, including phase recognition in many-body
systems~\cite{Carrasquilla2017}, neural-network representations of
quantum states~\cite{Carleo2017}, and learning-based quantum control
strategies~\cite{Bukov2018RL}.
These successes suggest that graph-based models may also be effective for guiding decision processes in variational quantum algorithms. Related graph-based models have also proven useful in chemically structured learning problems, including molecular excitation-property prediction, suggesting that message-passing architectures can capture nontrivial relationships between local quantum descriptors and global response behavior~\cite{Belaloui2025}. Recent work has further demonstrated that graph neural networks can
learn structured renormalization-group decimation rules in disordered quantum spin chains and reconstruct entanglement dynamics in monitored quantum circuits~\cite{vahedi2026a,vahedi2026b}. These studies highlight that physically meaningful hierarchical and spacetime structures can be captured by graph-based architectures. 
%\hl{In contrast, the present work focuses on accelerating adaptive
%variational quantum algorithms by learning gradient-based operator selection policies directly from Hamiltonian structure and state-dependent observables.} {\color{red} JV:: I think this is extra!!!}

While the present formulation is naturally motivated by disordered spin systems, the underlying selection problem is more general. In particular, adaptive variational algorithms in quantum chemistry face the same basic bottleneck: at each iteration, one must identify a small subset of operators from a large candidate pool that are most relevant for further improving the variational state. This suggests a broader learned-shortlisting strategy in which a machine-learning model does not necessarily replace the gradient oracle outright, but instead narrows the search to a tiny set of promising candidates before an exact local refinement step is applied. A central question is therefore whether the graph-based operator-selection framework developed here transfers, at least in proof-of-concept form, to small molecular active-space benchmarks of the type commonly used in adaptive-VQE and quantum-chemistry demonstrations~\cite{Claudino2020,Feniou2023,deGraciaTrivino2023}.

Training data are generated from exact simulations of disordered long-range spin chains, where the optimal operator at each step is identified using the commutator-based gradient magnitude
\begin{equation}
g_{ij} =
\left|
\langle \psi | [H,P_{ij}] | \psi \rangle
\right|,
\end{equation}
with $P_{ij}$ denoting two-site entangling generators. This representation naturally endows the operator pool with a graph structure, where nodes correspond to spins and edges correspond to candidate two-qubit operators.

The resulting model learns a policy that maps graph-structured physical information to operator-selection decisions. Once trained, the network predicts promising operators in a single forward pass, thereby avoiding the expensive gradient scan over the full operator pool required in standard ADAPT-VQE. We evaluate this approach on disordered long-range spin chains and compare the learned policy with gradient-based selection and heuristic strategies. Our results show that the GNN policy identifies relevant operators with substantially lower selection cost while maintaining competitive variational accuracy.

% \hl{More broadly, this work shows that adaptive circuit construction in variational quantum algorithms can be understood as a structured decision problem on interaction graphs. By combining ideas from variational quantum algorithms, disordered quantum systems, and graph-based machine learning, we introduce a framework for learning operator-selection policies that reduce classical computational overhead in adaptive quantum algorithms.}
In this work, we formulate adaptive operator selection as a graph-based sequential decision problem and develop a graph neural network that predicts the next entangling operation directly from the interaction graph and state-dependent observables. We train the model on exact simulations of disordered long-range spin chains and then test both its operator-ranking accuracy and its downstream variational performance.

\section{Learning Operator Selection in Disordered Model}

\subsection{Problem Setup and Variational Objective}
We consider disordered long-range quantum spin chains in which $N$ spin-$1/2$
degrees of freedom are distributed at random positions $\{r_i\}_{i=1}^N$
along a one-dimensional segment of length $L$ with open boundaries, 
as illustrated in the system-to-graph mapping shown in Fig.~\ref{fig:gnn_architecture_map}. In the dilute regime $n \ll 1$, the random spatial separations produce a broad distribution of couplings $J_{ij}$, resulting in strong bond disorder.

The interaction between spins $i$ and $j$ decays algebraically with
distance, $J_{ij}=J_0|r_i-r_j|^{-\alpha}$, where the exponent $\alpha$ controls the effective range of interactions.
Small values of $\alpha$ correspond to long-range interactions, while
large $\alpha$ approaches the short-range limit. This spatially disordered structure naturally defines an interaction graph
in which spins correspond to nodes and candidate two-qubit operators
$P_{ij}$ correspond to edges of the graph. This graph representation forms the basis
for the learning-based operator-selection strategy developed in this work. While our focus is on this class of disordered spin systems, the formulation introduced below applies more broadly to interacting quantum systems with graph-structured couplings.

The Hamiltonian is taken to be of XXZ form,
\begin{equation}
H = \sum_{i<j} J_{ij}
\left(
X_i X_j + Y_i Y_j + \Delta Z_i Z_j
\right),
\label{eq:H}
\end{equation}
which reduces to the Heisenberg model when $\Delta=1$.

Our objective is to approximate the ground state of $H$ using a variational
quantum circuit constructed from two-qubit entangling operations and local single-qubit rotations. Conventional hardware-efficient ans\"atze entangle
qubits according to fixed geometric patterns independent of the interaction
structure. In strongly disordered systems, however, the hierarchy of couplings
spans multiple energy scales, and entanglement develops in a highly
inhomogeneous and scale-dependent manner. Fixed entangling patterns may therefore fail to fully capture the intrinsic structure of the Hamiltonian.

We instead adopt a circuit-growth perspective. Starting from a product state
$|\psi_0\rangle$, we iteratively append entangling operations selected
adaptively based on the interaction graph. At depth $T$, the variational state
takes the form
\begin{equation}
|\psi_T(\boldsymbol{\theta})\rangle
=
\prod_{t=1}^{T}
U_{i_t j_t}(\theta_t)
\,|\psi_0\rangle,
\end{equation}
where each $U_{i_t j_t}$ acts on a selected pair of spins.

The variational objective is to minimize the energy
\begin{equation}
E(\boldsymbol{\theta})
=
\langle \psi_T(\boldsymbol{\theta})|H|\psi_T(\boldsymbol{\theta})\rangle.
\end{equation}

The central problem addressed in this work is therefore:
\emph{Given an interaction graph and an evolving variational state,
how can one efficiently select a sequence of entangling bonds $(i_t,j_t)$
that improves the variational energy?}

We formulate this as a sequential decision problem over the interaction graph
and explore whether a graph neural network can learn a bond-ranking policy
from graph-structured physical information.
\subsection{Graph Neural Network for Variational Circuit Construction}

Each Hamiltonian instance defined above is represented as a weighted
interaction graph $G=(V,E)$, where nodes correspond to spins and edges encode
effective couplings. Figure~\ref{fig:gnn_architecture_map} summarizes both the graph representation used for feature construction and the subsequent message-passing and edge-scoring workflow used to rank candidate operators. To ensure computational efficiency while preserving the
dominant energy scales, we retain for each node the $K$ strongest incident
couplings, yielding a sparsified but physically faithful graph representation
of the interaction structure.

\begin{figure*}[ht]
\centering
\includegraphics[width=\textwidth]{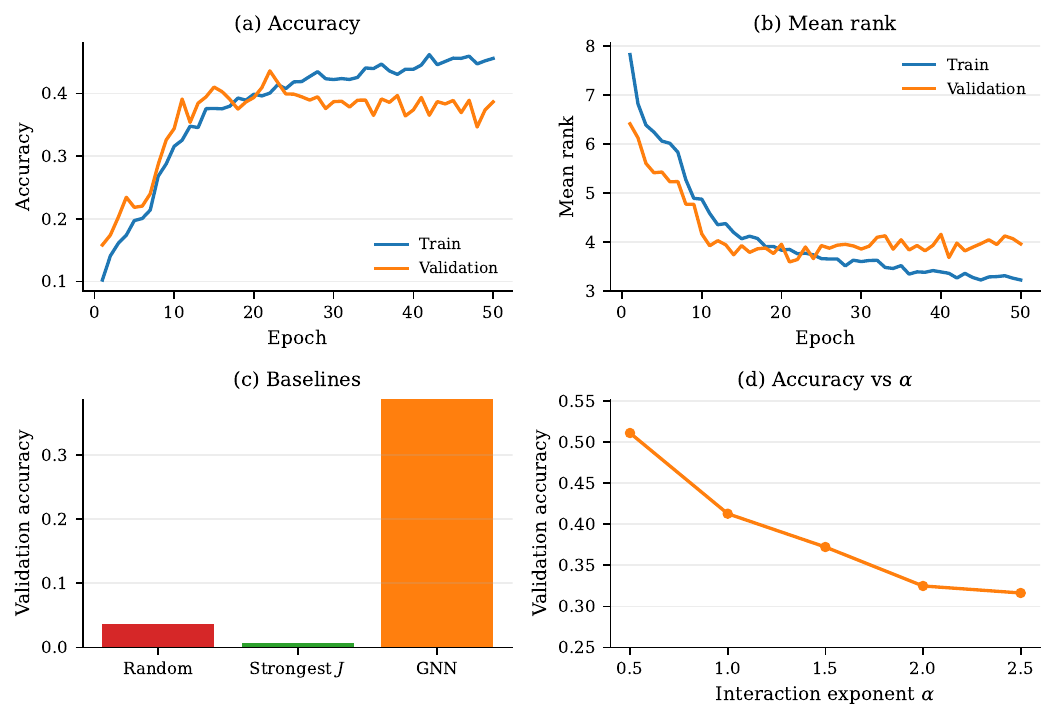}
\caption{
\textbf{Training dynamics and operator-selection performance of the learned GNN policy.}
(a) Training and validation accuracy during optimization of the pointer-based graph neural network. The model is trained to predict the operator with the largest gradient magnitude based on the interaction graph and state-dependent observables.
(b) Mean rank of the correct operator among all candidate edges during training. Lower values indicate better agreement with the greedy gradient oracle used to generate supervision labels.
(c) Comparison with simple baselines. The learned policy significantly outperforms both random edge selection and a heuristic that always selects the strongest coupling $J_{ij}$.
(d) Prediction accuracy as a function of the interaction-range exponent $\alpha$. The policy maintains consistent performance across different interaction regimes, demonstrating generalization from strongly long-range to more short-range interaction profiles.
}
\label{fig:training_summary}
\end{figure*}

\paragraph{Graph representation.}
Node features encode geometric and local interaction structure,
\begin{equation}
\mathbf{v}_i^{\text{static}} =
\bigl(
x_i,\,
d_i,\,
s_i,\,
m_i
\bigr),
\end{equation}
where $x_i=r_i/L$ denotes the normalized spatial coordinate, $d_i$ is the
nearest-neighbor spacing, $s_i=\sum_j J_{ij}$ is the total incident coupling
strength, and $m_i=\max_j J_{ij}$ is the strongest bond connected to spin $i$.
These quantities provide a compact summary of the local interaction landscape surrounding each spin.

Edge attributes capture both geometric and energetic structure. For each
retained edge we define
\begin{equation}
\mathbf{e}_{ij}^{\text{static}} =
\bigl(
\log |J_{ij}|,\,
\log |r_i-r_j|,\,
\rho_{ij}
\bigr),
\end{equation}
where $\rho_{ij}$ quantifies the relative dominance of the coupling $J_{ij}$
with respect to other interactions incident on its endpoints. Specifically,
we define
$\rho_{ij}=\tfrac{|J_{ij}|}{\max_{k} |J_{ik}|}+
\tfrac{|J_{ij}|}{\max_{k} |J_{jk}|}$. The logarithmic parametrization improves numerical stability in the
strong-disorder regime, where coupling strengths typically span multiple
orders of magnitude.

\paragraph{State-dependent features.}

In addition to the static graph structure, node and edge attributes are
augmented with expectation values computed with respect to the current
variational state $|\psi_t\rangle$. These observables encode the evolving
correlation structure of the wavefunction and allow the network to learn a
state-aware policy.

For nodes we include local magnetizations
\begin{equation}
\bigl(
\langle Z_i \rangle_t,\,
\langle X_i \rangle_t
\bigr),
\end{equation}
while candidate edges are assigned two-spin correlators
\begin{equation}
\bigl(
\langle Z_i Z_j \rangle_t,\,
\langle X_i X_j \rangle_t,\,
\langle Y_i Y_j \rangle_t
\bigr).
\end{equation}

The interaction decay exponent $\alpha$ is also provided as a global input
feature in order to permit generalization across different interaction ranges.

The resulting node and edge representations at step $t$ are therefore
\begin{align}
\mathbf{v}_i^{(t)} &=
\bigl(
\mathbf{v}_i^{\text{static}},
\langle Z_i \rangle_t,
\langle X_i \rangle_t,
\alpha
\bigr), \\
\mathbf{e}_{ij}^{(t)} &=
\bigl(
\mathbf{e}_{ij}^{\text{static}},
\langle Z_i Z_j \rangle_t,
\langle X_i X_j \rangle_t,
\langle Y_i Y_j \rangle_t,
\operatorname{Var}_{\psi_t}(P_{ij})
\bigr),
\end{align}
where the final component denotes the variance of the entangling generator
\begin{equation}
\operatorname{Var}_{\psi_t}(P_{ij}) =
\langle P_{ij}^2 \rangle_t -
\langle P_{ij} \rangle_t^2 .
\end{equation}
This quantity reflects the magnitude of quantum fluctuations associated with the operator $P_{ij}$ in the current state and provides information about its potential entangling effect.

\paragraph{Neural Network Architecture.}
The operator-ranking function $f_\theta$ is implemented using a
message-passing graph neural network designed to assign scores to
candidate edges.

Node features $\mathbf{v}_i$ and edge features $\mathbf{e}_{ij}$ are first
embedded into a latent representation space using linear encoders,
\begin{align}
\mathbf{h}_i^{(0)} &= \phi_n(\mathbf{v}_i), \\
\mathbf{e}_{ij}' &= \phi_e(\mathbf{e}_{ij}),
\end{align}
where $\phi_n$ and $\phi_e$ are multilayer perceptrons producing
hidden vectors of dimension $d$.

Information is then propagated through the graph using several
edge-conditioned message-passing layers. At layer $\ell$, messages are
constructed along edges according to
\begin{equation}
m_{ij}^{(\ell)} =
\phi_m\!\left(
\mathbf{h}_i^{(\ell)},\,
\mathbf{h}_j^{(\ell)},\,
\mathbf{e}_{ij}'
\right),
\end{equation}
where $\phi_m$ is a multilayer perceptron. Incoming messages are aggregated
using a permutation-invariant sum operation,
\begin{equation}
\mathbf{m}_i^{(\ell)} =
\sum_{j\in\mathcal{N}(i)} m_{ij}^{(\ell)} .
\end{equation}

Node embeddings are updated through
\begin{equation}
\tilde{\mathbf{h}}_i^{(\ell+1)} =
\phi_u\!\left(
\mathbf{h}_i^{(\ell)},\,
\mathbf{m}_i^{(\ell)}
\right),
\end{equation}
followed by a residual connection
\begin{equation}
\mathbf{h}_i^{(\ell+1)} =
\mathbf{h}_i^{(\ell)} +
\tilde{\mathbf{h}}_i^{(\ell+1)} .
\end{equation}
Several such message-passing iterations allow information about couplings,
correlations, and geometry to propagate across the interaction graph.

After the final message-passing layer, candidate edges are scored using
the embeddings of their endpoints together with encoded edge attributes. 
The scoring function takes the form
\begin{equation}
s_{ij} =
\phi_s\!\left(
\mathbf{h}_i,\,
\mathbf{h}_j,\,
\mathbf{h}_i-\mathbf{h}_j,\,
\mathbf{h}_i \odot \mathbf{h}_j,\,
\mathbf{e}_{ij}'
\right),
\end{equation}
where $\odot$ denotes element-wise multiplication and $\phi_s$ is a
multilayer perceptron producing a scalar output. These scores define the operator-ranking function used during circuit construction. As illustrated in Fig.~\ref{fig:gnn_architecture_map}, the edge-scoring head combines the final endpoint embeddings, their difference, their element-wise product, and the encoded edge feature to produce a scalar score for each candidate operator.

The graph neural network employs node and edge embeddings of dimension ($d=96$) and consists of five edge-conditioned message-passing layers. The interaction graph is sparsified by retaining the $K$ strongest couplings incident on each spin (see Dataset Generation for details). In the numerical experiments we use $K=6,7,8$ for system sizes $N=8,10,12$, respectively. During dataset generation, circuit construction proceeds using a small fixed rotation angle ($\theta_0=0.05$) when applying entangling operations. These choices were found to provide a good balance between expressivity and computational efficiency in the simulated systems considered here.

\paragraph{Sequential circuit construction.}

Starting from an initial product state $|\psi_0\rangle$, sampled randomly from
the Bloch sphere for each spin, the variational circuit is constructed
iteratively. At step $t$ the graph neural network assigns a scalar score to
every candidate edge,
\begin{equation}
s_{ij}^{(t)} =
f_\theta\!\left(
\mathbf{v}_i^{(t)},\,
\mathbf{v}_j^{(t)},\,
\mathbf{e}_{ij}^{(t)}
\right),
\end{equation}
where $f_\theta$ is implemented via message passing followed by an edge-scoring
head.

The edge with the largest score is selected,
\begin{equation}
(i_t,j_t) =
\arg\max_{(i,j)\in E} s_{ij}^{(t)},
\end{equation}
and an entangling operator is appended to the circuit,
\begin{equation}
U_{i_t j_t}(\theta_t)
=\exp\!\left(-i\theta_t P_{i_t j_t}\right)
\end{equation}
where $P_{i_t j_t}=\left(
X_{i_t}X_{j_t}
+
Y_{i_t}Y_{j_t}
+
\Delta Z_{i_t}Z_{j_t}\right).$ Unlike strong-disorder renormalization-group procedures that permanently remove
bonds from the system, edges remain available throughout circuit construction
and may be selected multiple times if beneficial for energy reduction. This allows the variational ansatz to incorporate loop structures and long-range correlations, rather than being restricted to tree-like growth patterns.

During dataset generation, gradients are evaluated at $\theta_t=0$ while the
state update is performed using a small fixed rotation angle $\theta_0$,
\begin{equation}
|\psi_t\rangle =
U_{i_t j_t}(\theta_0)\,
|\psi_{t-1}\rangle .
\end{equation}

This procedure produces a discrete trajectory in Hilbert space that
approximates a greedy coordinate-descent process in the space of available
entangling operators. All expectation values required for feature construction
and supervision signals are computed using classical statevector simulation
during dataset generation.

\paragraph{Supervision and training.}

Training labels are derived from the first-order energy response associated
with candidate entangling operators. For each intermediate state $|\psi_t\rangle$
we evaluate the gradient magnitude
\begin{equation}
g_{ij}^{(t)} =
\left|
i
\langle \psi_t |
[H,P_{ij}]
|\psi_t\rangle
\right|,
\end{equation}
where $P_{ij}$ denotes the generator of the corresponding entangling gate.

This quantity corresponds to the magnitude of the energy derivative
\begin{equation}
\frac{d}{d\theta}
\langle \psi_t(\theta) |
H
|\psi_t(\theta) \rangle
\Big|_{\theta=0},
\end{equation}
and therefore identifies the operator that produces the largest instantaneous
change in energy when applied infinitesimally. This gradient-based operator ranking is closely related to the selection rule used in adaptive ansatz construction methods such as ADAPT-VQE~\cite{Grimsley2019ADAPT,Tang2021QubitADAPT}. In ADAPT-VQE, the operator with the largest gradient magnitude is appended to the circuit at each iteration and the circuit parameters are subsequently reoptimized. In the present work, the gradient rule is used only during dataset generation to produce supervision signals, while operator selection at deployment time is performed directly by the learned GNN policy.

To avoid numerical instabilities arising from large variations in gradient
magnitude, the scores are normalized within each step according to
\begin{equation}
\tilde g_{ij}^{(t)}
=
\frac{g_{ij}^{(t)}}{\max_{(k,l)\in E} g_{kl}^{(t)} + \epsilon},
\end{equation}
where $\epsilon$ is a small constant.

A soft teacher distribution over candidate edges is then defined as
\begin{equation}
q_{ij}^{(t)}
=
\frac{
\exp(\tilde g_{ij}^{(t)}/\tau)
}{
\sum_{(k,l)\in E}
\exp(\tilde g_{kl}^{(t)}/\tau)
},
\end{equation}
with temperature parameter $\tau$ controlling label sharpness.

The GNN produces logits $s_{ij}^{(t)}$ that define the student distribution
\begin{equation}
p_{ij}^{(t)}
=
\frac{
\exp(s_{ij}^{(t)})
}{
\sum_{(k,l)\in E}
\exp(s_{kl}^{(t)})
}.
\end{equation}

Training minimizes the cross-entropy between teacher and student distributions,
\begin{equation}
\mathcal{L}
=
-\frac{1}{T}
\sum_{t=0}^{T-1}
\sum_{(i,j)\in E}
q_{ij}^{(t)}
\log p_{ij}^{(t)}.
\end{equation}

Dataset splits are performed at the level of disorder realizations in order to
prevent leakage of correlated circuit-construction trajectories between
training and evaluation sets.

\paragraph{Dataset Generation.}
Training data are generated using exact classical simulations of disordered
long-range spin chains defined by the Hamiltonian in Eq.~\ref{eq:H}. For each
realization, $N$ spins are placed at random integer positions
$r_i \in \{0,\dots,L\}$ along a one-dimensional segment, producing a spin
density $n=N/L$. The interaction strengths are determined from the
power-law form $J_{ij}=J_0/|r_i-r_j|^{\alpha}$, where the decay exponent
$\alpha$ controls the interaction range. 

To reduce the size of the candidate operator pool while preserving the
dominant energy scales, the interaction graph is sparsified by retaining,
for each spin, the $K$ strongest couplings. The resulting set of edges
defines the pool of candidate two-spin generators $P_{ij}$.

Starting from a random product state sampled independently on the Bloch sphere for each spin, we simulate short circuit-growth trajectories of length $T$. At each step, the commutator magnitude $g_{ij}=
\left|
i\langle\psi_t|[H,P_{ij}]|\psi_t\rangle
\right|$ is evaluated for all candidate edges, and the edge with the largest value is selected as the training label. The quantum state is then updated using
a small entangling rotation
$|\psi_{t+1}\rangle=
e^{-i\theta_0 P_{ij}}|\psi_t\rangle$, 
which generates a trajectory of intermediate states used to construct
state-dependent node and edge features.

Datasets are generated for several system sizes
$(N,L)=(8,80),(10,100),(12,120)$, corresponding to a fixed disorder density
$n=0.1$. For each system size and interaction exponent
$\alpha\in\{0.5,1.0,1.5,2.0,2.5\}$ we generate multiple disorder realizations,
and each trajectory step is stored as a separate graph sample for training
the neural network.

\paragraph{Inference and compiled ansatz.}
At inference time the trained graph neural network replaces the gradient
evaluation procedure and produces a bond-selection sequence tailored to each
disorder realization. The resulting sequence defines the topology of the
variational circuit.

All entangling angles are subsequently optimized using standard variational
quantum eigensolver (VQE) techniques. Because parameters are reoptimized
globally during this stage, the final variational state is not restricted to
the discrete trajectory used during supervision.

\paragraph{Connection to renormalization ideas.}
The circuit-growth procedure bears a qualitative resemblance to strong-disorder
renormalization-group ordering rules. In the early stages of circuit
construction, when the state remains weakly entangled, the gradient magnitude
$g_{ij}^{(t)}$ is typically dominated by the corresponding coupling strength
$J_{ij}$. In this regime the greedy rule approximately reduces to selecting
the strongest interaction, paralleling the ordering principle underlying SDRG~\cite{Fisher1994,stefan2025,stefan2026}.

As the circuit deepens and correlations develop, however, the gradient signal
becomes sensitive to many-body expectation values, causing the bond-selection
policy to deviate from purely local coupling strength. The learned policy
therefore interpolates between simple interaction-strength heuristics and a
fully state-aware operator selection rule that adapts to the evolving quantum
state.

\section{Numerical Results}

We now evaluate the performance of the proposed graph neural network (GNN) policy
for operator selection in adaptive variational circuit construction.
Unless stated otherwise, results are averaged over independent disorder
realizations of long-range spin chains defined by Eq.~\eqref{eq:H},
with fixed density $n=0.1$ and interaction exponents
$\alpha \in \{0.5,1.0,1.5,2.0,2.5\}$.
All energy errors are defined relative to the exact ground-state energy
obtained by exact diagonalization.
We distinguish between three complementary evaluation targets throughout this section.
First, we assess how accurately the GNN reproduces the gradient-based operator-ranking rule used to generate supervision labels.
Second, we evaluate the quality of the stepwise circuit-growth trajectories obtained when the next operator is selected either by the gradient oracle or by the learned policy under the same discrete rollout protocol.
Third, we test the quality of the final compiled ansatz after full variational reoptimization of all circuit parameters.
This distinction is important because agreement with the greedy one-step oracle, energy reduction along the rollout trajectory, and final post-reoptimization variational accuracy probe related but not identical aspects of performance.

\begin{figure}[htp]
\centering
\includegraphics[width=\columnwidth]{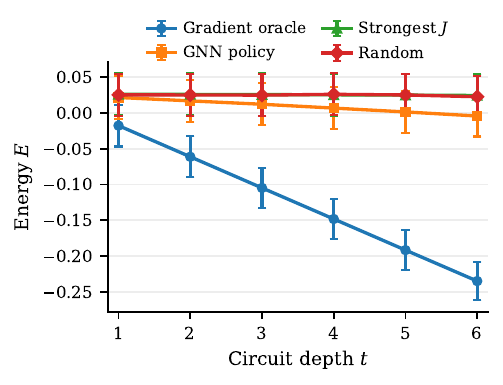}
\caption{ Energy expectation value as a function of circuit depth t across different operator-selection strategies, averaged over disordered long-range spin chain realizations. The trajectories represent a discrete rollout using the fixed small-angle update rule ($\theta$) from dataset generation prior to global parameter reoptimization. The learned GNN policy consistently outperforms both the random and static strongest-coupling baselines. A visible performance gap remains relative to the exact gradient oracle, which utilizes perfect local commutator information. Error bars denote the standard error across disorder realizations.
}
\label{fig:energy_depth}
\end{figure}

\paragraph{Learning the gradient-based selection rule}

Figure~\ref{fig:training_summary} summarizes the training behavior and
operator-ranking performance of the GNN policy.
Panel (a) shows the evolution of the top-1 prediction accuracy during
training. The model rapidly improves from near-random performance to a
validation accuracy of approximately $40\%$, while the training accuracy
reaches higher values, indicating a moderate generalization gap.
Given typical operator pool sizes $|E| \sim 20$--$30$, random selection
would yield accuracies of only a few percent, suggesting that the model captures nontrivial structure in the gradient-based selection rule.

A more informative metric is shown in Fig.~\ref{fig:training_summary}(b),
which reports the mean rank of the oracle-selected operator among all
candidate edges. While random ranking would place the correct operator
near the middle of the list, the trained model consistently assigns it
a mean rank of approximately four on the validation set. This indicates that even when the top prediction is incorrect, the oracle edge is typically placed among the highest-ranked candidates, suggesting that the learned policy captures key features relevant to the gradient magnitude.

Panel (c) compares the learned policy to simple baselines. Random edge selection gives an accuracy of only a few percent, consistent with the typical operator-pool size. The strongest-coupling heuristic performs poorly, indicating that the gradient-based operator label is not determined solely by the magnitude of the bare coupling ($J_{ij}$). This is expected because the commutator signal depends on the current quantum state and on the surrounding interaction environment, rather than only on the selected bond strength. In contrast, the GNN policy substantially outperforms both baselines, demonstrating that state-dependent observables and graph-structured information are essential for learning the operator-ranking rule.

Finally, Fig.~\ref{fig:training_summary}(d) shows the prediction accuracy as a function of the interaction exponent ($\alpha$). The policy performs best in the more long-range regimes and gradually decreases in accuracy as the system becomes more short-range dominated. Nevertheless, the learned policy remains well above random selection across all values of ($\alpha$), indicating that it captures transferable structure across interaction ranges.

Overall, these results indicate that the GNN policy is able to learn meaningful structure in the gradient-based operator selection rule. 
At the same time, imitation accuracy alone does not fully determine downstream variational performance:
a learned policy may deviate from the exact one-step ranking while still producing competitive or even favorable multi-step operator sequences.
For this reason, we next examine both the quality of the rollout trajectories generated by the policy and the quality of the final ansatz after full parameter reoptimization.

\paragraph{Energy reduction during circuit growth}

We next examine how the learned operator-selection policy affects the variational energy during circuit construction. Figure~\ref{fig:energy_depth} shows the energy expectation value as a function of circuit depth when operators are appended according to different selection strategies. Here the comparison is performed at the level of the discrete rollout trajectory itself: at each step, an operator is selected and applied using the same small fixed update rule as in dataset generation. Accordingly, Fig. \ref{fig:energy_depth} probes operator-selection quality under a common greedy construction protocol rather than the quality of the final fully reoptimized variational ansatz. The gradient oracle, which evaluates the exact signed commutator gradient
$i\langle\psi|[H,P_{ij}]|\psi\rangle$ at every step and applies the corresponding downhill update, provides a reference corresponding to greedy selection based on complete local gradient information. As expected, this oracle achieves the largest energy reduction throughout the circuit-growth process. The learned GNN policy consistently improves upon the simple graph-based heuristics based solely on interaction strengths, demonstrating that the network has learned nontrivial information about the operator-ranking structure beyond the bare couplings. However, a noticeable gap remains relative to the gradient oracle, indicating that the policy only partially captures the state-dependent information required to reproduce the exact ADAPT-VQE selection rule. Both the random policy and the strongest-coupling heuristic produce nearly flat energy trajectories, showing that interaction strength alone is a poor predictor of the operator that yields the largest instantaneous energy decrease. This observation is consistent with the training-set analysis, which revealed only weak correlation between the coupling magnitude and the commutator-gradient labels.

Overall, these results suggest that the learned GNN captures a meaningful fraction of the operator-selection signal and successfully identifies many of the important operators chosen by the oracle. At the same time, the remaining performance gap highlights the difficulty of reconstructing the full gradient landscape from local graph features alone and motivates future improvements in the policy architecture and training objective.

\begin{figure}[htp]
\centering
\includegraphics[width=\columnwidth]{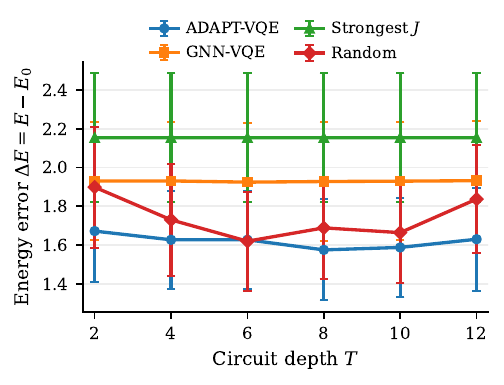}
\caption{
Energy error $\Delta E = E - E_0$ as a function of circuit depth $T$ for
different operator-selection strategies after full variational reoptimization of circuit parameters. This figure therefore probes the quality of the compiled ansatz topology induced by each selection strategy, rather than the quality of the intermediate rollout trajectory shown in Fig.~\ref{fig:energy_depth}. ADAPT-VQE achieves the lowest energy error by selecting operators using gradient information at each iteration. The GNN-VQE approach replaces this gradient evaluation with a learned policy based on graph-structured features. While the resulting circuits exhibit larger energy errors than those obtained with ADAPT-VQE,
the learned policy consistently improves upon the strongest-coupling heuristic, although random operator selection remains competitive for the system sizes considered here.}
\label{fig:depth_scaling}
\end{figure}

\paragraph{Full variational optimization}
To assess the quality of the operator sequences after full variational reoptimization of circuit parameters, we perform global VQE optimization over all entangling angles once the complete operator ordering has been determined. Figure~\ref{fig:depth_scaling} shows the resulting energy error $\Delta E = E - E_0$ as a function of the total circuit depth $T$.
This evaluation is fundamentally distinct from the stepwise rollout comparison in Figure~\ref{fig:energy_depth}. Here, the selected operator sequence is treated as a compiled, static ansatz topology, and all variational parameters are subsequently co-optimized globally. The resulting energy error therefore provides a direct measure of the final expressivity and state-preparation quality inherent to each operator-selection strategy.
As expected, ADAPT-VQE achieves the lowest energy error across all circuit depths, reflecting the clear structural advantage of utilizing exact gradient information at every single iteration. The GNN-VQE approach produces higher energy errors than ADAPT-VQE and remains relatively flat across depths, indicating that the imitation inaccuracies accumulated by the policy disrupt the global optimization landscape of the compiled circuit.
The static strongest-coupling heuristic performs worst among the structured strategies considered ($\Delta E \approx 2.15$), confirming that prioritizing bare interaction strengths alone leads to structurally restricted ansätze that stall even under full parameter reoptimization. While GNN-VQE successfully outperforms this static baseline—demonstrating that the integration of state-dependent and graph-structured information yields an improved basis for ansatz construction—an anomalous trend emerges regarding the random selection policy.
Interestingly, after full parameter optimization, the random policy (red curve) exhibits a notable drop in energy error, outperforming GNN-VQE across most circuit depths and briefly matching ADAPT-VQE at $T=6$. This behavior highlights a key distinction between greedy rollout performance and global parameter optimization in small system sizes. While a random sequence fails to minimize energy step-by-step with fixed angles, its arbitrary non-local gate placement can establish a highly expressive hardware-efficient ansatz layout. Once full multi-parameter optimization is permitted, this structural randomness can effectively bypass the localized subspaces or premature structural bottlenecks that a partially optimized GNN policy might inadvertently lock into.
Importantly, while ADAPT-VQE requires evaluating costly commutators over the entire operator pool at every single layer, the GNN-VQE policy bypasses this explicit scaling bottleneck with a single neural-network inference step. Although the resulting compiled circuits do not yet match the optimization performance achieved by the exact gradient-based procedure, the learned policy offers a computationally inexpensive strategy for informed operator selection, highlighting the potential of graph-based learning approaches for reducing the steep classical overhead of adaptive quantum circuit construction.

\begin{figure*}[ht] 
\centering 
\includegraphics[width=\textwidth]{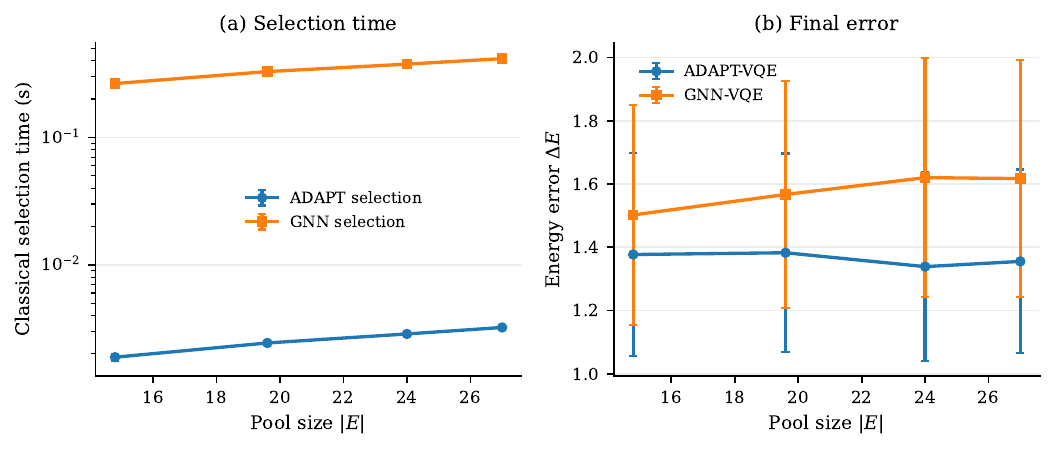} 
\caption{
Scaling of adaptive circuit construction with operator pool size $|E|$. 
\textbf{(a):} Time required to select $T$ operators during circuit construction. ADAPT-VQE evaluates the commutator gradient for every operator in the pool, leading to a selection cost that increases approximately linearly with $|E|$. The GNN-based policy predicts the next operator via a neural-network forward pass, replacing explicit gradient scans by classical inference. In the present small-scale classical simulations, however, the measured wall-clock selection time remains larger than ADAPT-VQE due to graph-construction and inference overhead, even though the qualitative dependence on $|E|$ is similar. \textbf{(b):} Energy error $\Delta E = E - E_0$ obtained after variational optimization of the resulting circuits. ADAPT-VQE achieves the lowest energy error, while the GNN policy produces circuits whose final variational accuracy remains competitive with simple heuristics, although it does not match ADAPT-VQE.}
\label{fig:pool_scaling} 
\end{figure*}

\paragraph{Scaling with operator pool size}
A central motivation for this framework is to mitigate the severe computational scaling associated with explicit gradient evaluation over large operator pools in adaptive variational algorithms. To clarify the precise computational tradeoffs, we compare the algorithmic complexity and measured execution cost of operator selection between the baseline ADAPT-VQE and our GNN-based approach.

In standard ADAPT-VQE, each structural iteration requires evaluating the energy gradient magnitude $g_{ij} = | i\langle\psi_t|[H,P_{ij}]|\psi_t\rangle |$ for every individual operator contained in the candidate pool. Letting $|E|$ denote the total pool size and $C_{\mathrm{exp}}$ represent the computational or hardware cost of evaluating a single commutator expectation value, the per-iteration selection cost scales strictly as
\begin{equation}
\mathcal{O}(|E|, C_{\mathrm{exp}}).
\end{equation}
Compounding this over a circuit of total depth $T$ yields an aggregate selection scaling of $\mathcal{O}(T |E| C_{\mathrm{exp}})$. On actual quantum hardware, $C_{\mathrm{exp}}$ is exceptionally punitive, requiring repeated state preparation and distinct quantum measurement scaling for every operator in the pool, creating a physical measurement overhead that grows linearly with pool size.

In contrast, the GNN-VQE policy replaces this extensive operator scan with a single forward pass through a message-passing neural network. For a network architecture utilizing an internal embedding dimension $d$ and operating with $L$ message-passing layers, the classical computational cost per layer scales as
\begin{equation}
\mathcal{O}(L |E| d^2),
\end{equation}
dictated primarily by the edge-conditioned latent updates. This leads to a total selection complexity over depth $T$ of $\mathcal{O}(T L |E| d^2)$. While both methodologies scale linearly with respect to the number of candidate edges $|E|$, their underlying physical expressions differ fundamentally. ADAPT-VQE requires active quantum state manipulation and destructive measurements for every single candidate operator, whereas the GNN architecture shifts this tracking entirely to classical tensor operations, ensuring a vastly lower hardware cost per candidate operator during practical scaling.

In the classical simulations reported here, the selection time corresponds to the wall-clock CPU time required to determine the optimal operator at each circuit step. For ADAPT-VQE, this includes the explicit, exact evaluation of the commutator gradient for all candidate operators using full classical statevector simulation. For the GNN policy, this time isolates the conversion of raw graph arrays into PyTorch geometric data structures combined with the single forward inference pass of the network. The reported inference benchmarks isolate the pure selection step given precomputed features; Table~\ref{tab:timing_breakdown} explicitly summarizes the exact operational breakdown included in these timing measurements.

Figure~\ref{fig:pool_scaling}(a) compares the measured classical operator-selection time required to construct circuits of fixed depth as a function of the pool size $|E|$ on a logarithmic scale. Both methods track a stable linear scaling trend with expanding pool sizes. However, within this small-scale classical verification regime, the GNN implementation exhibits a significantly higher wall-clock runtime overhead than ADAPT-VQE. This discrepancy is an expected artifact of classical benchmarking: computing a statevector commutator for a small matrix is computationally trivial on a CPU, meaning that the GNN's classical data-structure parsing and neural-network framework overhead dominate the raw execution time. Consequently, for small system sizes evaluated classically, the policy does not provide a practical runtime reduction.

\begin{table}[t]
\centering
\caption{Components included in the operator-selection timing measurements.}
\begin{tabular}{lcc}
\hline
Operation & ADAPT-VQE & GNN-VQE \\
\hline
Gradient evaluation over pool & $\checkmark$ & $\times$ \\
Graph feature assembly & $\times$ & $\checkmark$ \\
Neural-network inference & $\times$ & $\checkmark$ \\
State update after selection & $\times$ & $\times$ \\
Full VQE reoptimization & $\times$ & $\times$ \\
\hline
\end{tabular}
\label{tab:timing_breakdown}
\end{table}

Figure~\ref{fig:pool_scaling}(b) illustrates the corresponding final energy errors $\Delta E = E - E_0$ achieved after full variational parameter reoptimization of the resulting compiled circuits. Due to its reliance on perfect, non-local gradient calculation, ADAPT-VQE maintains a minimum, flat energy error ($\Delta E \approx 1.38$) across all pool sizes. GNN-VQE yields slightly larger errors that exhibit a mild upward drift from $\approx 1.50$ to $\approx 1.62$ as the pool expands to $|E|=27$. This minor performance degradation is expected, as a larger operator pool introduces a higher density of candidate edges, increasing the probability of top-1 selection errors from the policy. Crucially, the performance gap between the two methods remains tightly bounded and does not diverge rapidly, proving that the learned network reliably generalizes its ranking capabilities even when embedded in noisier graph environments.

Taken together with the scaling behavior in Figure~\ref{fig:depth_scaling}, these results demonstrate that the GNN policy successfully compiles highly competitive variational ansätze without ever directly querying the physical gradient landscape during deployment. While small-scale classical statevector emulations favor explicit calculations, the core advantage of our graph-based policy lies in substituting hardware-prohibitive quantum measurement scaling for constant-time classical neural inference—a distinction that is expected to become a defining factor as systems scale toward classically unsimulable regimes or deploy directly on physical quantum processors.

\begin{figure*}[t]
\centering
\includegraphics[width=\textwidth]{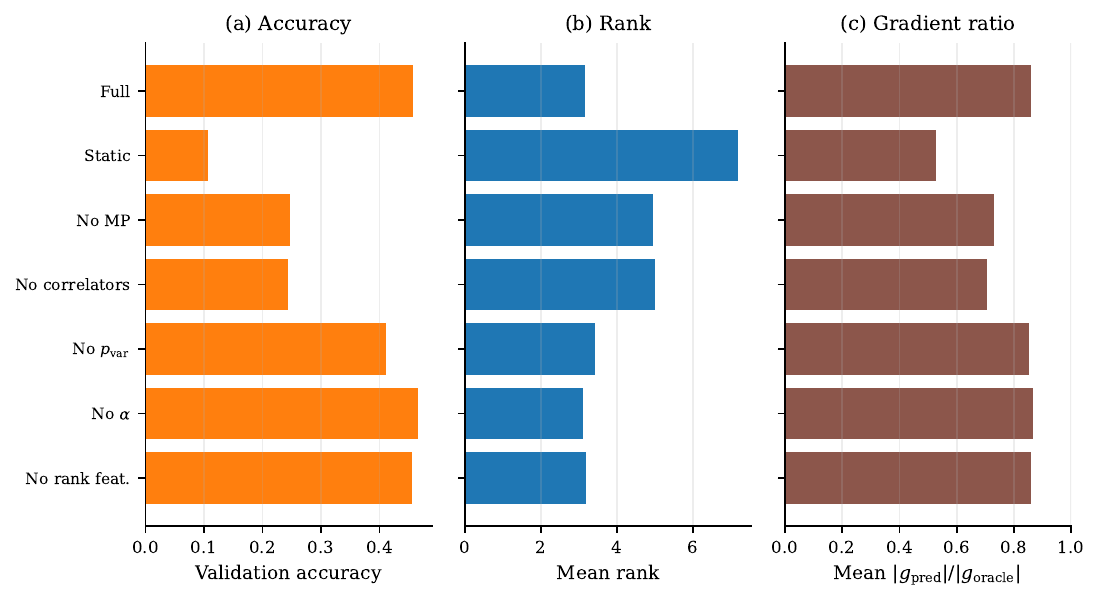}
\caption{
Ablation study of the proposed GNN operator-selection policy. \textbf{(a):} validation accuracy for predicting the oracle-selected operator. \textbf{(b):} mean rank of the correct operator (lower is better). \textbf{(c):} mean gradient-quality ratio
$\langle |g_{\mathrm{pred}}|/|g_{\mathrm{oracle}}| \rangle$.
The full model combines static graph features, state-dependent correlator information, operator-variance features, interaction exponent information, and graph message passing.
The largest performance degradation occurs when only static graph information is retained or when message passing and correlator features are removed, indicating that state-dependent information and relational message propagation are the primary contributors to performance.
In contrast, removing the interaction exponent $\alpha$ or the rank feature has negligible impact, while removing the operator-variance feature produces only a modest reduction in accuracy.
These results suggest that the predictive power of the model is driven predominantly by state-dependent correlator features together with graph message passing.
}
\label{fig:ablation}
\end{figure*}

\paragraph{Ablation study}

To identify which input features and architectural components contribute most strongly to the performance of the operator-selection policy, we performed a series of targeted ablations. Figure~\ref{fig:ablation} summarizes the validation accuracy, mean rank of the correct operator, and gradient-quality ratio
$\langle |g_{\mathrm{pred}}|/|g_{\mathrm{oracle}}| \rangle$
for several reduced variants of the model.

The strongest degradation occurs when only static graph information is retained. The \texttt{static\_only} model achieves a validation accuracy of only $0.107$, compared with $0.457$ for the full model, while the mean rank deteriorates from $3.16$ to $7.18$. This result demonstrates that state-dependent features are essential for predicting useful operator choices and that static interaction information alone is insufficient.

Removing message passing also substantially reduces performance, lowering the validation accuracy to $0.247$ and increasing the mean rank to $4.94$. A similar degradation is observed when correlator features are removed, yielding an accuracy of $0.244$ and mean rank $4.99$. These results indicate that both relational information propagation and state-dependent correlation features play a central role in the learned policy.

In contrast, removing the operator-variance feature $p_{\mathrm{var}}$ produces only a modest reduction in performance ($0.410$ accuracy and mean rank $3.41$), suggesting that this feature provides useful but non-essential information. Interestingly, removing the interaction exponent $\alpha$ or the operator-rank feature has almost no effect on performance. The \texttt{no\_alpha} and \texttt{no\_rank\_feature} variants achieve validation accuracies comparable to, or slightly exceeding, that of the full model, indicating that these quantities contribute little additional predictive power once the remaining graph and state features are available.

The gradient-quality metric exhibits the same qualitative trends. The full model achieves a mean gradient ratio of $0.858$, whereas the static-only variant drops to $0.526$. Models without message passing or correlator channels also show noticeable reductions ($0.729$ and $0.705$, respectively), confirming that these components improve the network's ability to identify operators associated with large energy gradients.

Overall, the ablation study indicates that the performance of the proposed architecture is driven primarily by the combination of state-dependent correlator information and graph message passing. Auxiliary features such as operator variance provide smaller gains, whereas the explicit inclusion of $\alpha$ and rank-based features appears largely unnecessary for the system sizes considered here.

\section{Transfer to molecular active-space benchmarks}
To assess whether the learned operator-selection framework extends beyond disordered spin chains, we constructed two small molecular active-space benchmarks based on LiH and BeH$_2$ in the STO-3G basis using Jordan--Wigner-mapped qubit Hamiltonians. Small active-space molecules of this type are widely used as controlled test cases in adaptive-VQE and quantum-computing-for-chemistry studies~\cite{Claudino2020,Feniou2023,Singh2022,Toth2020,Belaloui2025}.
In both cases, the adaptive-selection problem is formulated in direct analogy to the spin-model setting: a current variational state defines state-dependent features, a finite pool of candidate operators is scored using commutator-based oracle information during dataset generation, and learned models are trained to rank the next operator. The molecular benchmarks are not intended as large-scale quantum-chemistry demonstrations, but rather as controlled transfer tests of the learned-shortlist idea in a chemically meaningful setting. In this context, the use of reduced active spaces is deliberate: active-space construction is itself a nontrivial methodological choice in quantum chemistry and remains central to near-term quantum algorithms for molecular problems~\cite{deGraciaTrivino2023,Stein2016,King2021,Kolodzeiski2023}.

\begin{figure*}[t]
\centering
\includegraphics[width=\textwidth]{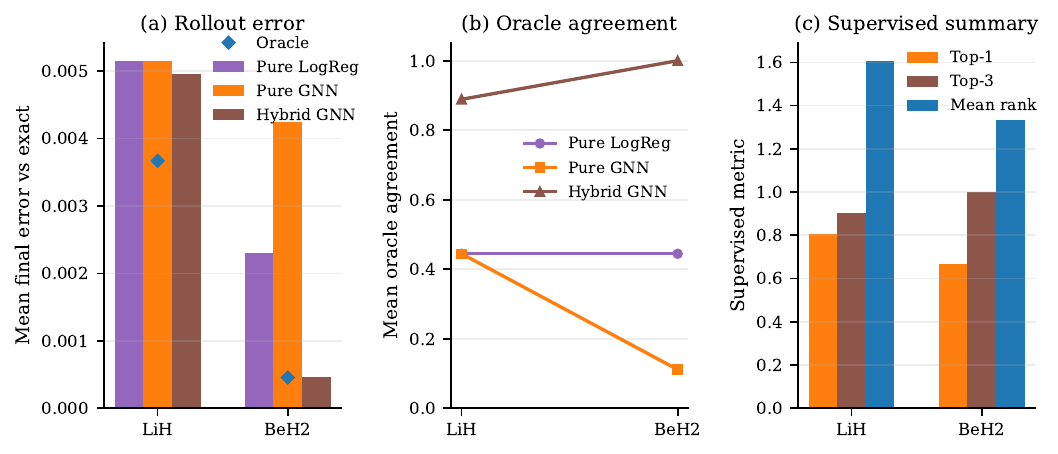}
\caption{
\textbf{Cross-benchmark transfer of the learned-shortlist workflow to molecular active-space problems.}
(\textbf{a}) Mean final rollout energy error relative to the exact ground state for LiH and BeH$_2$, comparing pure logistic-regression ranking, pure tuned GNN ranking, and the hybrid GNN policy with exact rescoring over a small shortlist.
(\textbf{b}) Mean oracle agreement during rollout for the same methods.
(\textbf{c}) Supervised ranking summary of the tuned soft-label GNN on the two molecular benchmarks, reported in terms of top-1 accuracy, top-3 accuracy, and mean true rank.
The main message is that pure learned policies are not uniformly reliable as standalone rollout strategies, but the learned GNN consistently contains useful ranking information. When used as a shortlist generator and combined with exact local rescoring, it recovers near-oracle behavior on LiH and oracle-level behavior on BeH$_2$.
}
\label{fig:molecular_transfer}
\end{figure*}

The cross-benchmark results are summarized in Fig.~\ref{fig:molecular_transfer}. For LiH, the best tuned soft-label GNN reaches a supervised top-1 accuracy of $0.804$, top-3 accuracy of $0.902$, and mean true rank of $1.608$, while the preferred hybrid shortlist size is $k=4$ out of a total candidate count of $14$. In rollout evaluation, the hybrid GNN improves the mean oracle agreement from $0.444$ for the pure learned policies to $0.889$ and reduces the final rollout error from $0.0052$ to $0.00496$. BeH$_2$ is particularly useful in this context because it has appeared repeatedly as a small but nontrivial benchmark in quantum-computing studies of molecular electronic structure, while still remaining compact enough for controlled active-space experiments~\cite{Singh2022,Toth2020,Sun2024}. For BeH$_2$, the supervised top-1 accuracy of the tuned GNN is $0.6667$, with top-3 accuracy $1.0000$ and mean true rank $1.3333$, but the more important result is the rollout behavior: with only $k=2$ shortlisted candidates out of $52$ total, hybrid rescoring recovers oracle-level rollout performance, with final error equal to the oracle reference and mean oracle agreement $1.0000$.

These molecular results clarify the practical role of the learned model. The strongest interpretation is not that the GNN already provides the best standalone policy in every setting. Indeed, on BeH$_2$ the pure GNN rollout is worse than the logistic baseline. Rather, the robust conclusion across both molecules is that the GNN learns a high-quality ranking over candidate operators, and that this ranking is most valuable when used to restrict the exact search to a very small subset of candidates. In this sense, the learned model acts as a \emph{screening stage} rather than as a complete replacement for the oracle.

% \begin{figure}[t]
% \centering
% \includegraphics[width=\columnwidth]{figures/figure_molecular_cost_quality.pdf}
% \caption{
% \textbf{Cost--quality tradeoff of hybrid shortlist rescoring on the molecular benchmarks.}
% The horizontal axis shows the fraction of the full candidate pool that must still be rescored exactly after GNN shortlisting, while the vertical axis shows the final rollout error relative to the exact ground state.
% For LiH, the preferred hybrid setting uses $k=4$ out of $14$ candidates, corresponding to a cost fraction of $0.2857$ and near-oracle performance.
% For BeH$_2$, the preferred hybrid setting uses only $k=2$ out of $52$ candidates, corresponding to a cost fraction of $0.0385$, yet recovers oracle-level rollout behavior.
% This indicates that the learned GNN can substantially reduce the expensive exact search while retaining the practical benefit of oracle-guided operator selection.
% }
% \label{fig:molecular_cost_tradeoff}
% \end{figure}

The cost--quality tradeoff is particularly clear when the preferred shortlist sizes are compared directly. On LiH, the hybrid workflow reaches its best observed performance using $4/14$ candidates, corresponding to a cost fraction of $0.286$. On BeH$_2$, the preferred hybrid workflow uses only $2/52$ candidates, corresponding to a cost fraction of $0.038$, while still reproducing the oracle rollout exactly. Taken together, these two benchmarks indicate that learned shortlisting is not merely a convenient reformulation of the selection problem, but can meaningfully reduce the exact search burden in chemically motivated active-space settings. From a chemistry perspective, this is consistent with the broader idea that machine learning may be most useful in near-term quantum workflows not as a complete replacement for established quantum-chemical structure, but as an intermediate layer that narrows expensive downstream searches or optimizations~\cite{Blunt2022,Alexeev2025}. Although these systems remain deliberately small, they provide a concrete proof-of-concept that the learned-shortlist strategy transfers beyond the original spin-model domain and may serve as a useful intermediate layer between full gradient scans and fully learned policies.

\section{Conclusion}

In this work we introduced a learning-based approach for operator
selection in adaptive variational quantum algorithms.
By reformulating the operator-selection step as a graph-based decision
problem, we developed a graph neural network (GNN) policy that predicts
the next entangling operator directly from the interaction graph and
state-dependent observables.
The learned policy is trained to approximate the gradient-based
selection rule used in adaptive ansatz construction methods such as
ADAPT-VQE.

Numerical experiments on disordered long-range spin chains indicate
that the GNN policy captures key structure in the greedy gradient-based
selection rule.
When used to construct variational circuits, the resulting GNN-VQE
approach achieves energy errors that remain close to those obtained with
gradient-based ADAPT-VQE, while reducing the computational overhead
associated with scanning the operator pool.
In particular, the learned policy replaces repeated gradient evaluations with a single neural-network inference step, thereby avoiding repeated full-pool gradient scans during operator selection while maintaining competitive variational performance.

The long-range spin models considered in this work are directly relevant
to several experimental quantum simulation platforms.
Trapped-ion systems naturally realize power-law interacting spin
Hamiltonians of the form $J_{ij} \propto |i-j|^{-\alpha}$ through
phonon-mediated interactions, with tunable interaction exponent
$\alpha$~\cite{Britton2012,Islam2013,Richerme2014,Jurcevic2014}.
Long-range spin interactions have also been engineered in Rydberg atom
arrays via dipole--dipole or van der Waals couplings~\cite{Browaeys2020,Adrian2022},
as well as in superconducting qubit processors with programmable
couplers and mediated interactions~\cite{Kjaergaard2020}.
These platforms provide promising opportunities for implementing
adaptive variational circuits and testing learning-assisted operator
selection strategies on near-term quantum hardware.

More broadly, our results suggest that adaptive circuit construction in
variational quantum algorithms exhibits structured patterns that can be
learned from graph-encoded physical information. The molecular
active-space benchmarks considered here further indicate that the most
robust role of the learned model is as a shortlist generator: rather
than directly replacing gradient-based selection, the GNN can identify
a tiny set of promising candidates that are then refined by exact local
rescoring. Across LiH and BeH$_2$, this hybrid learned-oracle workflow
recovers near-oracle or oracle-level rollout behavior while using only
a small fraction of the full candidate search. This points toward a
practical route for reducing classical overhead in adaptive
quantum-classical algorithms while preserving the physically meaningful
structure of operator selection. Future work could explore transfer of this strategy across broader Hamiltonian families, integration with hardware-efficient ansätze, and extension to larger many-body and molecular systems where explicit gradient evaluation becomes increasingly costly~\cite{Sun2025,Sun2024}.

% \acknowledgments
% We acknowledge the Ataro Group for generously providing office space and computational resources that facilitated the completion of this work.

% \appendix 

\bibliography{references/ref}
\end{document}